\documentclass[dvips,sii]{ipart}

\usepackage{graphics, graphicx, tikz}
\RequirePackage[OT1]{fontenc}
\RequirePackage{amsthm,amsmath, amssymb, tikz}
\RequirePackage[round]{natbib}
\RequirePackage[colorlinks]{hyperref}
\RequirePackage{hypernat}

\pubyear{2011}
\volume{0}
\issue{0}
\firstpage{1}
\lastpage{8}
\arxiv{math/0000.0000}

\startlocaldefs

\newtheorem{theorem}{Theorem}[section]

\endlocaldefs

\begin{document}

\begin{frontmatter}
\title{Spatial Multiresolution Cluster Detection Method}
\runtitle{Spatial MCD method}

\begin{aug}
\author{\fnms{Lingsong}
\snm{Zhang}\thanksref{t1}\ead[label=e1]{lingsong@purdue.edu}},
\address{Department of Statistics \\ Purdue University \\150 N. University St.\\
West Lafayette, IN, 47906\\
\printead{e1}}
\and
\author{\fnms{Zhengyuan} \snm{Zhu}\thanksref{t3}\ead[label=e2]{zhuz@iastate.edu}},
\address{Department of Statistics \& Statistical Laboratory\\ Iowa State University \\ Snedecor Hall \\ Ames IA 50011-1210\\ \printead{e2}}

\thankstext{t1}{Zhang's research is partially supported by Purdue Research Foundation International Travel Grant 51060311.}
\thankstext{t3}{Zhu's research is partially supported by NRCS-ISU cooperative agreement number 68-7482-11-534.}
\runauthor{Zhang and Zhu}

\affiliation{Purdue University and Iowa State Univiersity}
\end{aug}

\received{\sday{22} \smonth{11} \syear{2011}}

\begin{abstract}
A novel multi-resolution cluster detection (MCD) method is proposed to identify irregularly shaped clusters in space. Multi-scale test statistic on a single cell is derived based on likelihood ratio statistic for Bernoulli sequence, Poisson sequence and Normal sequence. A neighborhood variability measure is defined to select the optimal test threshold. The MCD method is compared with single scale testing methods controlling for false discovery rate and the spatial scan statistics using simulation and f-MRI data. The MCD method is shown to be more effective for discovering irregularly shaped clusters, and the implementation of this method does not require heavy computation, making it suitable for cluster detection for large spatial data.
\end{abstract}


\end{frontmatter}

\section{Introduction}
Spatial cluster detection is an important problem in many different fields, such as epidemiology and image analysis. In this paper, we focus on the
following spatial cluster detection problem: to detect an (irregular) spatial region which has a different signal intensity compared to the background. Specifically, we focus on a large spatial region, e.g. an image or a geographical map. We assume that when there is no signal in it, the data
collected at all locations are independent and follow the same distribution with an unknown parameter. Under the alternative hypothesis,
within an (unknown) sub-region, the data observed are from a different distribution. Let
$\mathcal{D}$ be the entire spatial region, $s\in \mathcal{D}$ be the location in $\mathcal{D}$, and $Y(s)$ be the observed data at location
$s$. Let $\mathcal{D}_\mathcal{A}$ be a subregion in $\mathcal{D}$, i.e., $\mathcal{D}_\mathcal{A} \subset \mathcal{D}$. We assume that $Y(s)
\stackrel{iid}{\sim} f(\theta_0)$ under $H_0$. Under $H_1$, $Y(s) \stackrel{iid}{\sim} f(\theta_1)$, when $s \in \mathcal{D}_\mathcal{A}$; and $Y(s)
\stackrel{iid}{\sim} f(\theta_0)$, when $s \in \mathcal{D}/\mathcal{D}_\mathcal{A}$. Typical examples include
\begin{enumerate}
\item Binomial distribution, $f(\theta_0)=Bin(N, p_0)$, and $f(\theta_1)=Bin(N, p_1)$, with $p_1>p_0$.
\item Poisson distribution, $f(\theta_0)=P(\lambda_0)$, and $f(\theta_1)=P(\lambda_1)$, with $\lambda_1>\lambda_0$.
\item Normal distribution, $f(\theta_0)=N(\mu_0, \sigma^2)$ and $f(\theta_1)=N(\mu_1, \sigma^2)$, with $\mu_1 >\mu_0$.
\end{enumerate}
Our main objective is to identify the unknown region $\mathcal{D}_\mathcal{A}$. Here we do not have any assumptions on the shapes of
$\mathcal{D}$ and $\mathcal{D}_\mathcal{A}$.

One popular method to identify such spatial cluster is the spatial scan statistics. Scan statistics was first developed in
\cite{naus1965distribution, naus1965clustering}. See comprehensive review in \cite{glaz1999scan, glaz2001scan, glaz2009scan}. Kulldorff extended scan
statistics into multidimensional case which includes spatial problems \citep{kulldorff1999spatial}. See also in \cite{kulldorff2006elliptic,
costa2009applications, Loh:Lind:Wage:resi:2008, Zhan:Lin:clus:2009}. Kulldorff also developed a software to apply spatial scan statistics to spatial temporal data
\citep{kulldorff2010satscan}. The basic idea of scan statistics is to use scan windows of different sizes and locations and perform likelihood ratio tests on all the scan windows using simulation. Typical
window shapes in \cite{kulldorff2010satscan} include circle or ellipse. Other methods for detecting spatial clusters include modified Knox test \citep{Bake:modi:2004}, point process method \citep{Diggle:et.al:2005}, Bayesian hierarchical method \citep{Lian:Bane:Carl:baye:2009}, methods based on $K$ function \citep{wheeler2007, Loh2011}.
The spatial scan statistics methodology is shown to be very powerful compared to some other spatial cluster detection methods \citep{Kull:Tang:Park:powe:2003}. However, the power of the test is reduced when the shape of the true cluster is not circle or ellipsoid. It is also less useful when part of the inference objective is to identify the shape of the clusters.

Another commonly used approach is to perform test at each location in $\mathcal{D}$, and use multiple comparison methods to overcome the issue of large number of tests. One typical multiple comparison method is to control false discovery rate (FDR) \citep{benjamini1995controlling, storey2002direct}.
This method overcomes the fixed shape problem of the scan statistics, since it performs tests at every single individual location. However, it ignores the spatial information and the detected region is usually not spatially continuous. Recently this method has been extended by using an adjusted local false discovery rate method \citep{zhang2011multiple} for spatial clustering detection.

The multi-resolution cluster detection (MCD) method we propose in this paper takes advantage of the spatial information to make the test more powerful, while still maintains the flexibility of detecting irregularly shaped spatial clusters. The MCD method is motivated by the scale-space inference ideas in functional estimation and image analysis \citep{lindeberg1993scale, lindeberg1994scale,
chaudhuri1999sizer, chaudhuri2000scale}. \cite{zhang:zhu:marron:2007} and \cite{zhang2008multi} applied scale-space inference method to time series analysis with applications in Internet anomaly detection . The MCD method uses a similar idea as the scale-space inference by forming a test statistic based on multi-scale windows to effectively use the spatial information. A novel variability measure is used to identify the test threshold, which helps to maintain a balance between the sensitivity and specificity. The use of the variability measure overcomes the multiple comparison issues of the individual test, since it is a global procedure. Simulation studies and an application to fMRI data show that the MCD method compared favorably to the spatial scan statistic method and the multiple testing method based on FDR for identifying spatial clusters.

The remaining part of this paper is organized as follows. In Section \ref{methodsec} we describe the methodology and theory, which includes detailed derivation of the MCD method for binomial, Poisson and Normal distributions. The validity of the local variability measure is also shown in this section. Section \ref{simusec} provides simulation studies to compare different approaches for Binomial distributions with different alternative probability of success and different signal regions. An application of the three methods to a real fMRI data is given in Section \ref{examplesec}, and in Section \ref{discusssec} we discuss possible extensions.

\section{Methodology}  \label{methodsec}
Let $Y(s)$, $s\in \mathcal{D}$, be a sequence of independent random variables from the distribution $\mathcal{F}_{\theta(s)}$. Here $s$ is the location
in a two dimensional region $\mathcal{D}$, and $\theta(s)$ are the population parameters. In the later sections, if $s=(k, l)$, we may use $Y(s)$ and $Y_{kl}$ interchangeably. Under the null hypothesis, we assume $\theta(s)=\theta_0$ for all
$s \in \mathcal{D}$. Under the alternative hypothesis, we assume $\theta(s)=\theta_1$ for all $s \in \mathcal{D}_\mathcal{A} \subset \mathcal{D}$, and
$\theta=\theta_0$ otherwise.
Our objective is to identify $\mathcal{D}_\mathcal{A}$.
 We consider three examples in this paper: 1) $\mathcal{F}$ is $Bin(N_0, p)$, and $\theta=p$, the probability of success; 2) $\mathcal{F}$ is $Poisson(\lambda)$, and
$\theta=\lambda$; 3) $\mathcal{F}$ is $N(\mu, \sigma_0^2)$, and $\theta=\mu$. In the above three examples, $N_0$ and $\sigma_0^2$ are assumed to be known.

Our multi-resolution cluster detection (MCD) method has two steps: 1) At each location, we form a MCD test statistic which is expected to be large in the signal region, and small in the background region.  2) Determine a threshold for the test at each location which maintains a balance between specificity and sensitivity of the procedure. In Section \ref{MCDTestStatSec}, we describe the MCD test statistics and give detailed formula to compute the test statistics for the distributions mentioned above. In Section \ref{TestThresholdSec} we introduce the notion of neighborhood variability and describe a method to determine the detection threshold based on neighborhood variability. Some theoretical results are derived in Section \ref{validtestsec} to justify our method for setting the threshold. In Section \ref{noscalesec} we briefly discuss the choice of scales.

\subsection{The MCD test statistic} \label{MCDTestStatSec}
The test statistic for MCD at each location is formed using ideas from scale-space inference. In this subsection, let us focus on a specific location $s$. For easy
presentation, let $\mathcal{D}$ be a two dimensional grid with $n_r$ rows and $n_c$ columns, and $s=(i, j)$, with $i=1, 2, \ldots, n_r$, and $j=1, 2, \ldots, n_c$. Define a sequence of local regions $\{s\} = \mathcal{D}_1 \subset \mathcal{D}_2 \subset \cdots \subset \mathcal{D}_M \subset\mathcal{D}$. Note that our method does not
require $\mathcal{D}$ to be square, and the shapes of $\mathcal{D}_i$ can be irregular as well. Let us define the following aggregation vector
$(X_{ij}^1, X_{ij}^2, \cdots, X_{ij}^M)^T$, where $X_{ij}^1=Y_{ij}$, and
\begin{equation}
X_{ij}^r=\sum_{kl \in \mathcal{D}_r} Y_{kl}, \quad \textnormal{for all } 1\leq r \leq M.
\end{equation}

In this section, since we focus on a single location, we further simplify our notation $X_{ij}^r$ to be $X_r$.  Thus, the aggregation
vector is $(X_1, X_2, \cdots, X_M)^T$. Assuming that $(x_1, x_2, \cdots, x_M)^T$ is the observed vector, the likelihood ratio test statistics can be written as
\begin{align*}
\Lambda&=\frac{\sup_{\Theta_0}L(\theta; x)}{\sup_{\Theta}L(\theta; x)}\\
&=\frac{\sup P(X_1=x_1, X_2=x_2, \cdots, X_M=x_M|\Theta_0)}{\sup P(X_1=x_1, X_2=x_2, \cdots, X_M=x_M|\Theta)}.
\end{align*}
Based on standard theory on likelihood ratio test, under appropriate regularity conditions, we have
\[
-2\log (\Lambda) \sim \chi^2(M).
\]

Since $\mathcal{D}_\mathcal{A}$ is unknown, it is difficult to compute the denominator of the likelihood ratio statistic under the original alternative hypothesis. Instead we derive the formula for $\Lambda$ under the alternative that the parameter $\theta$ is a constant within each of the regions $\mathcal{D}_1$, $\mathcal{D}_2-\mathcal{D}_1$, $\cdots$, and $\mathcal{D}_k-\mathcal{D}_{k-1}$ for three commonly used distributions, Binomial, Poisson and Normal, and use them as the test statistics. The results does not depend on the shapes of $\mathcal{D}_r$. Thus, the user can use their favorite shape in the MCD method. We provide a function in R to calculate the test statistic, which provides two regular shapes (square and circle) in the implementation. In this paper, we will use the square shape for illustration purpose.

All the calculation next uses the following fact:
\begin{align*}
&P(X_1=x_1,\cdots, X_M=x_M)\\
&=P(X_1=x_1, X_2-X_1=x_2-x_1,
\cdots, \\
&\qquad X_M-X_{M-1}=x_M-x_{M-1})
\end{align*}
Note that since $Y_{ij}$ are independent, the $\{X_i-X_{i-1}\}$ are independent to each other. Using this formula, it is relatively straightforward to derive the results for three different distributions. In what follows we further introduce a notation $m_r =|\mathcal{D}_r|$, the cardinality of $\mathcal{D}_r$. We leave the details of the computation to the appendix.

\subsubsection{Binomial distribution}
Let $Y_{ij} \stackrel{iid}{\sim} Bin(N_{ij}, p_0)$ under the null hypothesis. Under the alternative, $Y_{ij} \stackrel{iid}{\sim} Bin(N_{ij}, p_k)$ for
all $(i, j) \in \mathcal{D}_k \backslash \mathcal{D}_{k-1}$, with $p_k > p_0$, and $Bin(N_{ij}, p_0)$ otherwise. It is relatively easy to get that, under the null hypothesis,
\[
X_r=\sum_{kl \in \mathcal{D}_r} Y_{kl} \sim Bin(N_r, p_0),
\]
where $N_r=\sum_{kl \in \mathcal{D}_r} N_{kl}$. It is straightforward that $N_i \leq N_{i+1}$. Let $\widehat{p}_{ij}=\frac{Y_{ij}+1}{N_{ij}+2}$ (note
that this adjustment is to avoid 0 and 1 as an estimate, which is similar to the Wilson's estimate). Define the follows:
\begin{align*}
\widehat{p}_0 &=\mathrm{Median}_{\mathcal{D}}(\widehat{p}_{ij}),\\
\widehat{p}_1 &=\max (\mathrm{Median}_{\mathcal{D}_1}(\widehat{p}_{ij}), \widehat{p}_0), \\
\widehat{p}_r &=\max (\mathrm{Median}_{\mathcal{D}_r\backslash \mathcal{D}_{r-1}}(\widehat{p}_{ij}), \widehat{p}_0).
\end{align*}
Here the maximization is to make sure the estimate under the alternative is not smaller than the estimate under the null hypothesis, and we use median instead of mean for robustness. Then we have the
following formulae to compute the approximated likelihood ratio test statistic for Binomial distribution:

\begin{align*}
-2&\log(\Lambda)=-2\log\left(\frac{\sup_{\Theta_0}L(\theta; x)}{\sup_{\Theta}L(\theta; x)}\right)\\
&\approx -2\left[x_1\left(\log(\widehat{p}_0)-\log(\widehat{p}_1)\right)+ (N_1-x_1) \left(\log(1-\widehat{p}_0)\right.\right.\\
&\quad \left.-\log(1-\widehat{p}_1)\right)+(x_2-x_1)\times(\log(\widehat{p}_0)-\log(\widehat{p}_2))+\\
& \qquad((N_2-N_1)-(x_2-x_1)) (\log(1-\widehat{p}_0)-\log(1-\widehat{p}_2)) \\
& \qquad +\cdots+ (x_k-x_{k-1})(\log(\widehat{p}_0)-\log(\widehat{p}_k))\\
& \qquad+ ((N_k-N_{k-1})-(x_k-x_{k-1}))\\
&\qquad \left.\times(\log(1-\widehat{p}_0)-\log(1-\widehat{p}_k))\right].
\end{align*}

\subsubsection{Poisson distribution}
Let $Y_{ij} \sim P(\lambda_0)$ under the null hypothesis, and $Y_{ij} \sim P(\lambda_k)$ for $(i, j)\in \mathcal{D}_k \backslash \mathcal{D}_{k-1}$, with $\lambda_k > \lambda_0$. It is clear that, under the null hypothesis,
\[
X_{ij}^k \sim P(m_k\lambda_0).
\]

Let $\widehat{\lambda}_0=\mathrm{Median}_\mathcal{D}(Y_{ij})$, and
\[
\widehat{\lambda}_i=\max\left(\frac{x_i-x_{i-1}}{m_i-m_{i-1}}, \widehat{\lambda}_0\right),
\]
then we have the following formulae to compute approximate likelihood ratio test statistic for Poisson distribution:

\begin{align*}
-2\log(\Lambda) &\approx -2[x_1(\log(\widehat{\lambda}_0)-\log(\widehat{\lambda}_1))+(\widehat{\lambda}_1-\widehat{\lambda}_0)\\
&\quad +(x_2-x_1)(\log(\widehat{\lambda}_0)-\log(\widehat{\lambda}_2))+m_2(\widehat{\lambda}_2-\widehat{\lambda}_0)] \\
&=-2[(x_k-x_{k-1})(\log(\widehat{\lambda}_0)-\log(\widehat{\lambda}_k))\\
&\quad+m_k(\widehat{\lambda}_k-\widehat{\lambda}_0)]
\end{align*}

\subsubsection{Normal distribution}
 We assume that under the null hypothesis, $Y_{ij} \sim N(\mu, \sigma^2)$ and they are independent. Under the alternative, $Y_{ij}
\sim N(\mu_k, \sigma^2)$ when $(i, j)\in \mathcal{D}_k \backslash \mathcal{D}_{k-1}$, and $\mu_k>\mu$.

Here we assume that $\sigma^2$ is given. For real application, $\sigma^2$ is unknown, thus, we will use a robust estimate for the variance.

Let $\widehat{\mu}_0=\mathrm{Median}_\mathcal{D}(Y_{ij})$, and
\[
\widehat{\mu}_i=\max\left(\frac{x_i-x_{i-1}}{m_i-m_{i-1}}, \widehat{\mu}_0\right).
\]
We have

\begin{align*}
-2\log (\Lambda)& \approx \frac{2x_1(\widehat{\mu}_1-\widehat{\mu}_0)}{\sigma^2}+\frac{\widehat{\mu}_0^2-\widehat{\mu}_1^2}{\sigma^2}\\
&\qquad+\frac{2(x_2-x_1)(\widehat{\mu}_2-\widehat{\mu}_0)}{\sigma^2}+\frac{m_2(\widehat{\mu}_0^2-\widehat{\mu}_2^2)}{\sigma^2}\\
&\qquad+\cdots+\frac{2(x_k-x_{k-1})(\widehat{\mu}_k-\widehat{\mu}_0)}{\sigma^2}\\
&\qquad+\frac{m_k(\widehat{\mu}_0^2-\widehat{\mu}_k^2)}{\sigma^2} \\
\end{align*}

\subsection{Test threshold} \label{TestThresholdSec}
The likelihood ratio test theory suggests that $\chi^2$ is the approximated distribution of the test statistic, and one way to determine the test
threshold is to control for FDR based on this asymptotic distribution. However, such procedure turns out to have poor performance due to the fact
that 1) the asymptotic distribution may not be a good approximation for small sample sizes; 2) the test statistics at different locations are
correlated.

\begin{figure}
\begin{center}
\begin{tikzpicture}
\filldraw[fill=black] (0, 0) circle (1mm);
\filldraw[fill=gray!50] (1, 0) circle (1mm);
\filldraw[fill=gray!50] (-1, 0) circle (1mm);
\filldraw[fill=gray!50] (0, 1) circle (1mm);
\filldraw[fill=gray!50] (0, -1) circle (1mm);
\draw (2, 0) circle (1mm);
\draw (-2, 0) circle (1mm);
\draw (0, 2) circle (1mm);
\draw (0, -2) circle (1mm);
\draw (-2, -2) circle (1mm);
\draw (-2, -1) circle (1mm);
\draw (-2, 1) circle (1mm);
\draw (-2, 2) circle (1mm);
\draw (-1, -2) circle (1mm);
\draw (-1, -1) circle (1mm);
\draw (-1, 1) circle (1mm);
\draw (-1, 2) circle (1mm);
\draw (1, -2) circle (1mm);
\draw (1, -1) circle (1mm);
\draw (1, 1) circle (1mm);
\draw (1, 2) circle (1mm);
\draw (2, -2) circle (1mm);
\draw (2, -1) circle (1mm);
\draw (2, 1) circle (1mm);
\draw (2, 2) circle (1mm);
\node at (0.3, -0.3) {\small $(i, j)$};
\draw [dashed] (-1.5, 0) -- (0, 1.5);
\draw [dashed] (-1.5, 0) -- (0, -1.5);
\draw [dashed] (0, -1.5) -- (1.5, 0);
\draw [dashed] (0, 1.5) -- (1.5, 0);
\end{tikzpicture}
\end{center}
\caption{The diagram for the neighborhood variability. At each location $(i, j)$, the variance of the four nearest cells plus the $(i, j)$ point will be calculated as the local variability. This helps to identify the edge of the detection boundary.} \label{VarDiagram}
\end{figure}
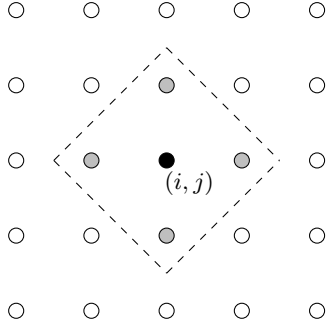

In this section, we introduce a neighborhood variability measure to determine the test threshold. Let $(i, j)$ be the location to calculate the local
variability. We choose the four nearest cells, where the $(i, j)$ cell serves as the center in the region, as shown in Figure \ref{VarDiagram}. The variance of these five data points is defined as the neighborhood variability.

Our assumption is that both signal region and the noise region have homogenous distribution within themselves. Thus, the local variability within these
two regions should be small, while the local variability for points close to the cluster boundary
should be relatively large.
In order to identify the ``optimal'' detection threshold, we compute the average neighborhood variability on the region with test statistics between two consecutive detection thresholds for
a sequence of monotone increasing test thresholds $t_1, t_2, \ldots, t_M$. When the threshold is too small, most of the signal regions are identified
as signal, along with many locations in the noise region. As the threshold increases, only the noise region are excluded, leading to a small average
neighborhood variability. Similarly when the threshold is too large, the average neighborhood variability between two consecutive detection regions will
also be small as they only include signal regions. We identify the two thresholds between which the average neighborhood variability is the highest and
take the average of these two as the optimal detection threshold.

To summarize, our MCD method can be carried out as follows:

\begin{enumerate}
\item At each location $s$, compute the MCD test statistic $T(s)$ at this location as described in Section 2.1 and the neighborhood variability measure $V(s)$
as described in Section 2.2.
\item Let $t_1=\min_s T(s)$, $t_M=\max_s T(s)$, and define an arithmetic sequence $t_1, t_2, \ldots, t_M$.

\item Let $\mathcal{D}_k=\{s: T(s)>t_k\}$,
and $V_k=\mbox{ave} \{V(s): s \in \mathcal{D}_k, s \notin \mathcal{D}_{k+1}\}$. Let
$k^*=\mbox{argmax}_k V_k $. The optimal test threshold is defined as $t^*=0.5(t_{k^*}+t_{k^*+1})$, and the final detected region is
defined as $\widehat{\mathcal{D}}_\mathcal{A}=\{s: T(s)>t^*\}$.

\end{enumerate}

\subsection{Validity of the test threshold} \label{validtestsec}
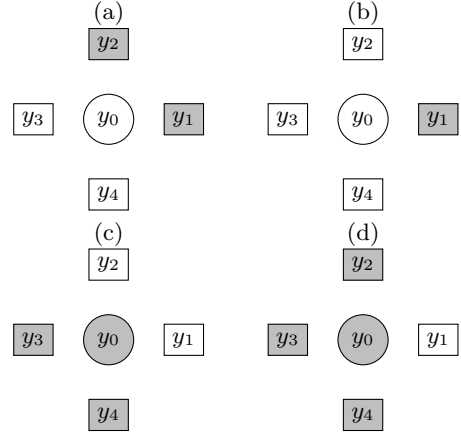
\begin{figure}[here]
\begin{tabular}{ccc}
(a) & \qquad & (b)  \\
\begin{tikzpicture}
\node at (0, 0) [circle, draw=black] {$y_0$};
\node at (1, 0) [rectangle, draw=black, fill=gray!50] {$y_1$};
\node at (0, 1) [rectangle, draw=black, fill=gray!50] {$y_2$};
\node at (-1, 0) [rectangle, draw=black] {$y_3$};
\node at (0, -1) [rectangle, draw=black] {$y_4$};
\end{tikzpicture}
&\qquad &
\begin{tikzpicture}
\node at (0, 0) [circle, draw=black] {$y_0$};
\node at (1, 0) [rectangle, draw=black, fill=gray!50] {$y_1$};
\node at (0, 1) [rectangle, draw=black] {$y_2$};
\node at (-1, 0) [rectangle, draw=black] {$y_3$};
\node at (0, -1) [rectangle, draw=black] {$y_4$};
\end{tikzpicture}
\\
(c) & \qquad &(d) \\
\begin{tikzpicture}
\node at (0, 0) [circle, draw=black, fill=gray!50] {$y_0$};
\node at (1, 0) [rectangle, draw=black] {$y_1$};
\node at (0, 1) [rectangle, draw=black] {$y_2$};
\node at (-1, 0) [rectangle, draw=black, fill=gray!50] {$y_3$};
\node at (0, -1) [rectangle, draw=black, fill=gray!50] {$y_4$};
\end{tikzpicture}
&\qquad &
\begin{tikzpicture}
\node at (0, 0) [circle, draw=black, fill=gray!50] {$y_0$};
\node at (1, 0) [rectangle, draw=black] {$y_1$};
\node at (0, 1) [rectangle, draw=black, fill=gray!50] {$y_2$};
\node at (-1, 0) [rectangle, draw=black, fill=gray!50] {$y_3$};
\node at (0, -1) [rectangle, draw=black, fill=gray!50] {$y_4$};
\end{tikzpicture}
\end{tabular}
\caption{\small The diagram for four types of boundary points. The white cells are noise points, and the gray cells are signal points. Panel (a) shows a boundary point at the noise region with an angle; panel (b) shows a boundary point at the noise region without an angle; panel (c) shows a boundary point at the signal region with an angle; and panel (d) shows a boundary point at the signal region without an angle.} \label{BoundaryPoint}
\end{figure}

In this subsection, we illustrate that under certain conditions, the MCD method can identify the signal region almost surely. Let $\mathcal{B}$ be the set of the boundary points, $\mathcal{N}$ be the set of inner points in the noise region, and $\mathcal{S}$ be the set of inner points in the signal region. For simplicity, we also use $\mathcal{B}^c=\mathcal{N}\cup \mathcal{S}$. Furthermore, let $n_b$ be the number of boundary points, $n_s$ be the number of (non-boundary) signal points, and $n_n$ be the number of (non-boundary) noise points. Let $N=n_b+n_n+n_s$ be the total number of data points in the spatial region $\mathcal{D}$. A boundary point is defined as the point where its local five observations (including the point itself) are not from the same group. See Figure \ref{BoundaryPoint} for illustrations of some boundary points.

For easy presentation, all the properties are shown under Normality assumption. We assume for any $Y$ in the noise region, $Y\sim N(0, 1)$; and for any $Y$ in the signal region, $Y\sim N(\delta, 1)$, where $\delta>0$. The parameter $\delta$ measures the signal strength. In addition, the proportion of boundary point $n_b/N$ is denoted as $p_B$. The following two theorems give some justification for the procedure we outlined in Section \ref{TestThresholdSec}. The proofs of both theorems are in the appendix.

\begin{theorem} The average test statistic $T$ at $\mathcal{N}$, $\mathcal{B}$ and $\mathcal{S}$ has the following relation: if $N, n_b, n_n, n_s \to \infty$, and $\delta>0$,
\[
P\left(\mathop{\mathrm{Ave}}_{Y\in \mathcal{N}} T(Y)<\mathop{\mathrm{Ave}}_{Y\in \mathcal{B}} T(Y)<\mathop{\mathrm{Ave}}_{Y\in \mathcal{S}} T(Y)\right)\to 1.
\] \label{AverageT}
\end{theorem}

\begin{theorem}
If  $\frac{\delta^2Np_B(1-p_B)}{8+4(1-p_B)} \to \infty$, then
\[
P\left(\mathop{\mathrm{Ave}}_{Y\in \mathcal{B}}{V}(Y)-\mathop{\mathrm{Ave}}_{Y\in \mathcal{B}^c}{V}(Y) >0\right) \to 1
\] \label{AverageV}
\end{theorem}

\begin{figure}
\begin{center}
\includegraphics[width=3.5in, angle=270]{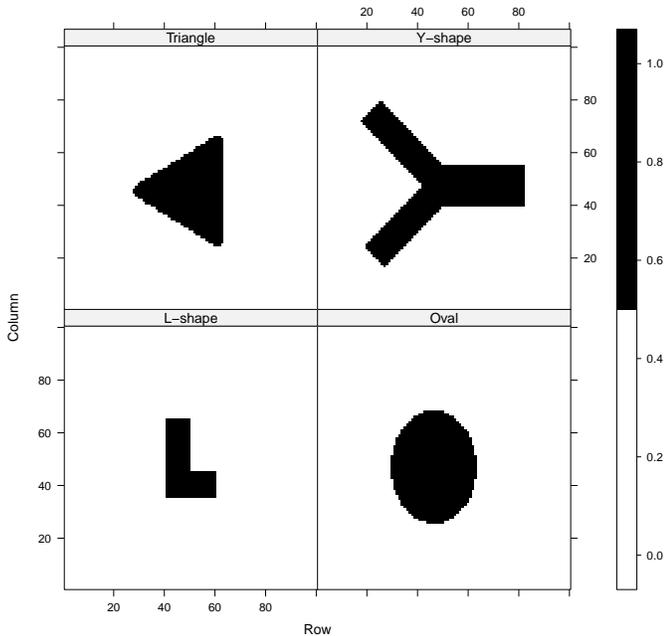}
\end{center}
\caption{\small The simulation settings} \label{SimuSetDiag}
\end{figure}

\noindent {\em Remark}: Theorem \ref{AverageT} shows that the MCD test statistic at points in the boundary region are on average smaller than those in the purely signal region and larger than those in the noise region. Thus which an appropriate sequence of test thresholds a substantial part of the boundary region can be in one of the belts defined by $\{s: s \in \mathcal{D}_k, s \notin \mathcal{D}_{k+1}\}$. Theorem \ref{AverageV} shows that the average variability measure on the boundary set is the largest if either the signal strength is large or the points in both boundary and non-boundary regions are large. Thus by selecting the belt with the maximum average variability, we have a good chance of identifying the true boundary of the signal region.

\subsection{Number of Scales} \label{noscalesec}
To use the MCD method, one need to determine: 1) how many scales to be used, and
2) what is the largest scale to be used.  The answer to these two questions are related to each other and depends on specific applications.
Typically, when the maximum scale is the same, the test power of the MCD method with more scales tends to be larger, as shown in one dimensional situation in
\cite{zhang:zhu:marron:2007}. For the two dimensional case, we compared two-scale vs. five-scale and single-scale methods and found that even though the test power of the five-scale method is better than the other two methods, the improvement from two to five is not as significant as those from one to two. Thus, in this paper, we will only use a two-scale MCD method for all the simulation examples and real application.

By aggregating local values, the MCD method will have some dilution effect, i.e., the noise points which are close
to the signal region are more likely to be identified as false signals. The larger the aggregation region, the higher chance of false detection for
these points. Thus, we will not use a very large window as the maximum scale. For a typical $100\times 100$ image, we find that the largest aggregation scale
around 10 (or a radius as 5) works well. In the following simulation and real examples, we will use window size 10 as our maximum scale.

\section{Simulation and Comparison} \label{simusec}
In this section, we use simulation to compare the MCD method with the single scale test method controlling FDR and the spatial scan statistics. All the examples are
simulated in a $100\times100$ map. To save space, in this manuscript, we only simulate binomial distribution. Poisson and Normal examples are provided
in the author's personal website. We use sensitivity and specificity to compare the performance of different methods. Under each simulation setting, we
will repeat the same simulation 100 times, and also report the probability of identification at each spatial location.

\begin{figure}
\begin{center}
\includegraphics[angle=270, width=4in]{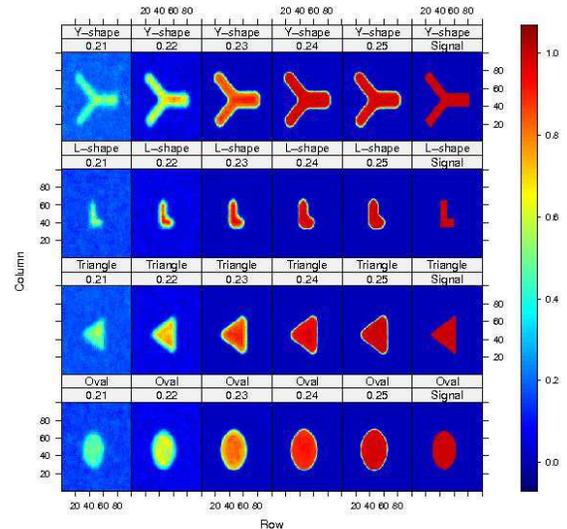}
\end{center}
\caption{The probability map of identification from the MCD method by different alternatives} \label{IdentifyFreqFig}
\end{figure}

\subsection{Simulation Settings}

Figure \ref{SimuSetDiag} shows the shapes of the simulations in this paper. We will study four different shapes of regions: $L$-shape, triangle, oval
and $Y$-shape. Note that the triangle and oval are convex, while $Y$ and $L$-shapes are not. Note that a data image contains
10,000 pixels. Table \ref{TruePixelCount} provides the number and the percentage of the pixels in the simulated signal regions.

\begin{table}[here]
\caption{Total true number of signal pixels} \label{TruePixelCount}
\begin{center}
\begin{tabular}{l|cccc}\hline
Shapes & $L$-shape & Oval & Triangle & $Y$-shape \\ \hline
Number of pixels & 400 & 1142 & 864 & 1344\\
(percentage) & 4\% & 11.42\% & 8.64\% & 13.44\% \\ \hline
\end{tabular}
\end{center}
\end{table}

For each simulation run, the background of the image is simulated as independent identically distributed Binomial distribution $Y(s) \sim Bin(100, 0.2)$,
if $s \in \mathcal{D}/\mathcal{D}_s$. The signal part, $Y(s) \sim Bin(100, p_1)$, if $s\in \mathcal{D}_s$, where $p_1 > 0.2$. In this paper, we focus
on five different alternative probabilities of success, $p_1=0.21$, 0.22, 0.23, 0.24, and 0.25. Note that when $p_1=0.21$, the detection should be very
challenging. We also assume that the population is preknown. For real applications such as epidemiology, the population of a specific geographical
location can be approximated by a recent survey. So this assumption is reasonable even for real applications.

\begin{figure}
\begin{center}
\includegraphics[angle=270, width=4in]{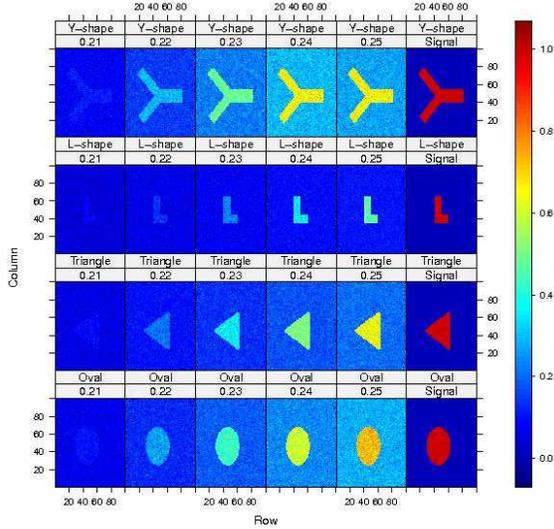}
\end{center}
\caption{The probability map of identification from the individual tests adjusted by false discovery rate by different alternatives}
\label{SingleScaleIdentifyFreqFig}
\end{figure}

\subsection{Simulation Results}
For each simulation setting, i.e., the same signal shape and the same alternative probability of success, we repeat the simulation 100 times. We use
the average of several measures to compare the performance of our method with other related methods. The measurements include specificity (the percentage of correctly identification of noise points), sensitivity (the percentage of correctly identification of signal points),
and the probability (relative frequency) of identifications at each pixels.

Tables \ref{AveRocTable} - \ref{AveRocTable4KScan} provides the average specificities and the average sensitivities for these different simulation
settings. The numbers in the parenthesis provides the standard deviation of the performance measures. It shows that when the alternative probability
increases, both sensitivity and specificity of the MCD method increases. The variability also decreases as the alternative probability increases. Note
that even when the alternative probability of success is 0.21, for all the shapes, our MCD method has an average sensitivity of 40\%, which is quite
impressive, although the variability is large. Note that when the alternative probability of success increases to 0.22, the specificity increases
dramatically to over 90\%, meanwhile, the sensitivity also increases.

\begin{figure}
\begin{center}
\includegraphics[angle=270, width=4in]{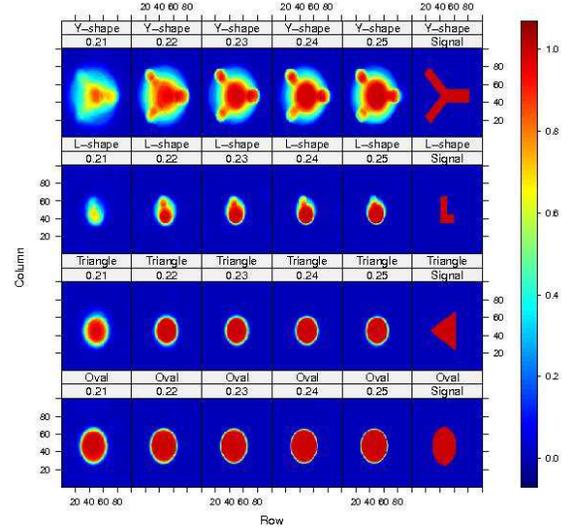}
\end{center}
\caption{The probability map of identification from the spatial scan method by different alternatives} \label{KScanIdentifyFreqFig}
\end{figure}

\begin{table*}[ht]
\caption{The average specificity/sensitivity for the MCD method by different simulation settings} \label{AveRocTable}
\begin{center}
\begin{tabular}{r|cc|cc|cc|cc}
  \hline
\quad & \multicolumn{2}{c|}{$L$-shape} & \multicolumn{2}{c|}{Oval shape} & \multicolumn{2}{c|}{Triangle shape} & \multicolumn{2}{c}{$Y$-shape} \\
\cline{2-9}
P1 & Specificity & Sensitivity & Specificity & Sensitivity& Specificity & Sensitivity& Specificity & Sensitivity \\
 \quad &  (Std) &  (Std)  &  (Std) &  (Std) &  (Std) &  (Std) &  (Std) &  (Std)\\
  \hline
0.21 & 0.8415 & 0.3818 & 0.8462 & 0.3972 & 0.8273 & 0.4036 & 0.8072 & 0.406 \\
   & (0.2899) & (0.3081) & (0.2865) & (0.3353) & (0.3142) & (0.323) & (0.3193) & (0.3558) \\ \hline
0.22 & 0.9401 & 0.6252 & 0.9309 & 0.5125 & 0.939 & 0.5806 & 0.9367 & 0.5299 \\
   & (0.1516) & (0.2574) & (0.1621) & (0.3865) & (0.1459) & (0.3533) & (0.1436) & (0.3832) \\ \hline
 0.23 & 0.9845 & 0.7986 & 0.9738 & 0.7669 & 0.9801 & 0.8079 & 0.9626 & 0.8232 \\
   & (0.0327) & (0.2358) & (0.0315) & (0.3744) & (0.0172) & (0.3159) & (0.0237) & (0.3313) \\ \hline
 0.24 & 0.987 & 0.9387 & 0.9769 & 0.9003 & 0.9774 & 0.9455 & 0.9484 & 0.98 \\
   & (0.0043) & (0.0999) & (0.0108) & (0.2776) & (0.0079) & (0.1827) & (0.0963) & (0.0911) \\ \hline
 0.25 & 0.9856 & 0.9723 & 0.9745 & 0.9817 & 0.9759 & 0.9923 & 0.96 & 0.9588 \\
   & (0.0032) & (0.0198) & (0.0061) & (0.1261) & (0.0043) & (0.0069) & (0.0118) & (0.1309) \\
\hline
\end{tabular}
\end{center}
\end{table*}


\begin{table*}[ht]
\caption{The average specificity/sensitivity for single scale detection method by different simulation settings} \label{AveSingleTable}
\begin{center}
\begin{tabular}{r|cc|cc|cc|cc}
  \hline
\quad & \multicolumn{2}{c|}{$L$-shape} & \multicolumn{2}{c|}{Oval shape} & \multicolumn{2}{c|}{Triangle shape} & \multicolumn{2}{c}{$Y$-shape} \\
\cline{2-9}
P1 & Specificity & Sensitivity & Specificity & Sensitivity& Specificity & Sensitivity& Specificity & Sensitivity \\
 \quad &  (Std) &  (Std)  &  (Std) &  (Std) &  (Std) &  (Std) &  (Std) &  (Std)\\
  \hline
0.21 & 0.9386 & 0.1003 & 0.9575 & 0.0709 & 0.9475 & 0.0845 & 0.9274 & 0.1148 \\
& (0.0203) & (0.0317) & (0.0164) & (0.0276) & (0.0207) & (0.0297) & (0.0262) & (0.0359) \\ \hline
0.22 & 0.8726 & 0.2613 & 0.9428 & 0.1382 & 0.9027 & 0.212 & 0.8532 & 0.2904 \\
& (0.0333) & (0.0536) & (0.0218) & (0.046) & (0.0266) & (0.0458) & (0.0348) & (0.0539) \\ \hline
0.23 & 0.8167 & 0.4322 & 0.9305 & 0.2277 & 0.8653 & 0.3556 & 0.7787 & 0.4854 \\
& (0.0427) & (0.0621) & (0.0265) & (0.0612) & (0.0323) & (0.0566) & (0.0459) & (0.0614) \\ \hline
0.24 & 0.7632 & 0.5976 & 0.9126 & 0.353 & 0.826 & 0.5122 & 0.7099 & 0.6568 \\
& (0.0451) & (0.0566) & (0.0222) & (0.0593) & (0.0386) & (0.0598) & (0.0597) & (0.0772) \\ \hline
0.25 & 0.7208 & 0.7272 & 0.8989 & 0.4755 & 0.7954 & 0.6464 & 0.7342 & 0.6584 \\
& (0.0434) & (0.0437) & (0.0238) & (0.0551) & (0.0367) & (0.0462) & (0.1309) & (0.2286) \\
   \hline
\end{tabular}
\end{center}
\end{table*}


\begin{table*}[ht]
\caption{The average specificity/sensitivity for the spatial scan method by different simulation settings} \label{AveRocTable4KScan}
\begin{center}
\begin{tabular}{r|cc|cc|cc|cc}
  \hline
\quad & \multicolumn{2}{c|}{$L$-shape} & \multicolumn{2}{c|}{Oval shape} & \multicolumn{2}{c|}{Triangle shape} & \multicolumn{2}{c}{$Y$-shape} \\
\cline{2-9}
P1 & Specificity & Sensitivity & Specificity & Sensitivity& Specificity & Sensitivity& Specificity & Sensitivity \\
 \quad &  (Std) &  (Std)  &  (Std) &  (Std) &  (Std) &  (Std) &  (Std) &  (Std)\\
  \hline
0.21 & 0.9819 & 0.4658 & 0.9809 & 0.8713 & 0.9738 & 0.734 & 0.9012 & 0.5461\\
 & (0.0238) & (0.2977) & (0.0283) & (0.0851) & (0.0245) & (0.1682) & (0.0996) & (0.2354)\\
0.22 & 0.9746 & 0.792 & 0.9885 & 0.9087 & 0.9831 & 0.7869 & 0.8697 & 0.791\\
 & (0.019) & (0.132) & (0.0071) & (0.0509) & (0.0136) & (0.0704) & (0.0863) & (0.1106)\\
0.23 & 0.9755 & 0.84 & 0.9898 & 0.9173 & 0.9855 & 0.7841 & 0.8788 & 0.8342\\
 & (0.0142) & (0.0802) & (0.0052) & (0.0347) & (0.0082) & (0.054) & (0.0782) & (0.0838)\\
0.24 & 0.9783 & 0.8486 & 0.9904 & 0.9192 & 0.9865 & 0.785 & 0.8762 & 0.8356\\
 & (0.0109) & (0.0832) & (0.0038) & (0.0278) & (0.0073) & (0.0478) & (0.0808) & (0.0792)\\
0.25 & 0.9779 & 0.8457 & 0.9902 & 0.9249 & 0.9865 & 0.7874 & 0.8677 & 0.8471\\
 & (0.0082) & (0.068) & (0.0035) & (0.0244) & (0.0072) & (0.0492) & (0.0793) & (0.0697)\\
\hline
\end{tabular}
\end{center}
\end{table*}

Table \ref{AveSingleTable} shows the specificity and sensitivity of the single scale method with test threshold selected to control false discovery rate. We use the direct false discovery rate method in \cite{storey2002direct}. In order to have a sensible detection result, we use FDR level as 0.60, which is relatively high.
Lower FDR level will lead to much smaller sensitivity. It shows that the specificity will decrease when the sensitivity increases. When the alternative probability is 0.22, the specificity of this method is similar to that of the MCD method, but the sensitivity is much smaller, which shows better performance of the MCD method over the single scale method.

Table \ref{AveRocTable4KScan} shows the specificity and sensitivity for the Spatial Scan method. It shows that when the data is oval or triangle, the
performance is among the best of these three methods, even when the signal strength is 0.21. But for $L$-shape and $Y$-shape, the sensitivity is
comparable to the MCD method. When the signal strength increases, the MCD method performs similarly as the spatial scan method, even for the two convex
shapes. The MCD method simultaneously have better specificity and sensitivity than the Spatial Scan method on the other two complicated shapes.

Figures \ref{IdentifyFreqFig} -\ref{KScanIdentifyFreqFig} provide the probability map of the pixels being detected by each of these three methods. The
probability is calculated by the relatively frequency of each pixels being detected by each method under the 100 runs under each simulation setting.
All figures shows that when the signal is weak, the chance of the signal region being detected is still much higher than the noise region.

Figure \ref{IdentifyFreqFig} provides the probability map by the MCD method. It shows that the actually shapes are almost correctly identified.
There is a slight dilution effect by showing oval shapes on the corners. But overall the complicated shapes such as $L$ and $Y$ are correctly
identified.

Figure \ref{SingleScaleIdentifyFreqFig} provides the probability map of the pixels being detected by the single scale method after adjusting for the false
discovery rate. On average, the shapes are correctly identified. But this figure shows that the probability of correctly identified is relatively
small. Even with a strong signal of 0.25, the probability of identification is about 60\%.

Figure \ref{KScanIdentifyFreqFig} provides a similar probability map of the pixels being detected by the Spatial Scan algorithm. We use the default circles
as the scan window. It shows that for the oval shape signal, even when the signal strength is low, the Spatial Scan algorithm provides excellent performance. For
other shapes, the performance is not bad in terms of the sensitivity. But it clearly misidentifies the shapes for triangle, $L$ and $Y$-shapes.

\subsection{Comparisons between different numbers of scales}
\begin{figure}[here]
\begin{center}
\includegraphics[angle=270,width=4in]{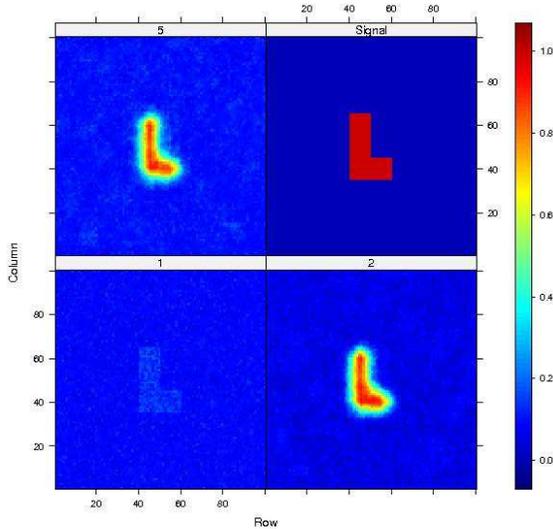}
\end{center}
\caption{\small Comparison of the detection probability for single scale, two-scale, and five-scale MCD methods, under the setting of $Bin(100, .2)$ vs. $Bin(100, .22)$} \label{MultiScaleComparePlot}
\end{figure}

Two-scale MCD method has been used in the previous subsection. In this subsection, we compare it with a five-scale MCD method, where the maximum scanning windows of these two methods are the same. In addition, we compare with another single scale method: the original scale tests using the local variability measure adjusted threshold. Here we only consider a single setting with the $L$-shape, with the alternative probability of success as 0.22. Figure \ref{MultiScaleComparePlot} shows the detection probability at each pixels.

\begin{figure}[here]
\begin{center}
\includegraphics[angle=270,width=3in]{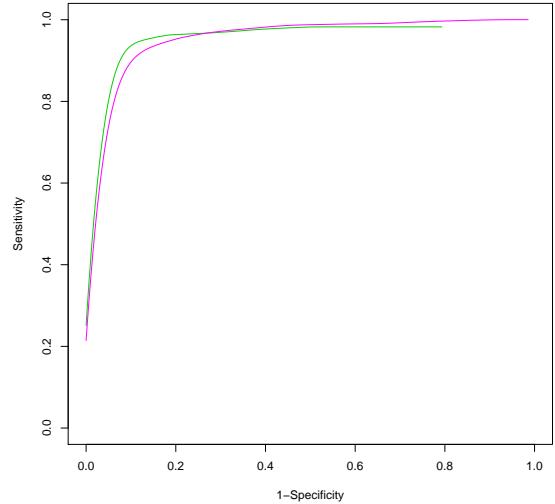}
\end{center}
\caption{\small The ROC curves for one specific simulation example. The green one is for a two-scale MCD method, and the purple one is for a five-scale MCD method} \label{Comp2vs5}
\end{figure}

It shows that the signal scale method plus the local variability measure adjustment is not powerful. In the signal region, it only has 20\% chance being detected. The two-scale MCD method with maximum scan window as 5 increases the detection probability to 80\%. The five-scale MCD method with the same maximum scan window is very similar to the two-scale one. We also compared the ROC curves between 2-scale and 5-scale. It turns out that on average the 5-scale ROC has larger area under the curve than the 2-scale method, but they are very close to each other. See one example ROC curve comparions in Figure \ref{Comp2vs5}. We note that the maximum scanning window is more important than the number of scales.

\begin{figure*}
\begin{center}
\begin{tabular}{cc}
(a) Raw Data & (b) Spatial Scan method \\
\includegraphics[angle=270, width=3.2in]{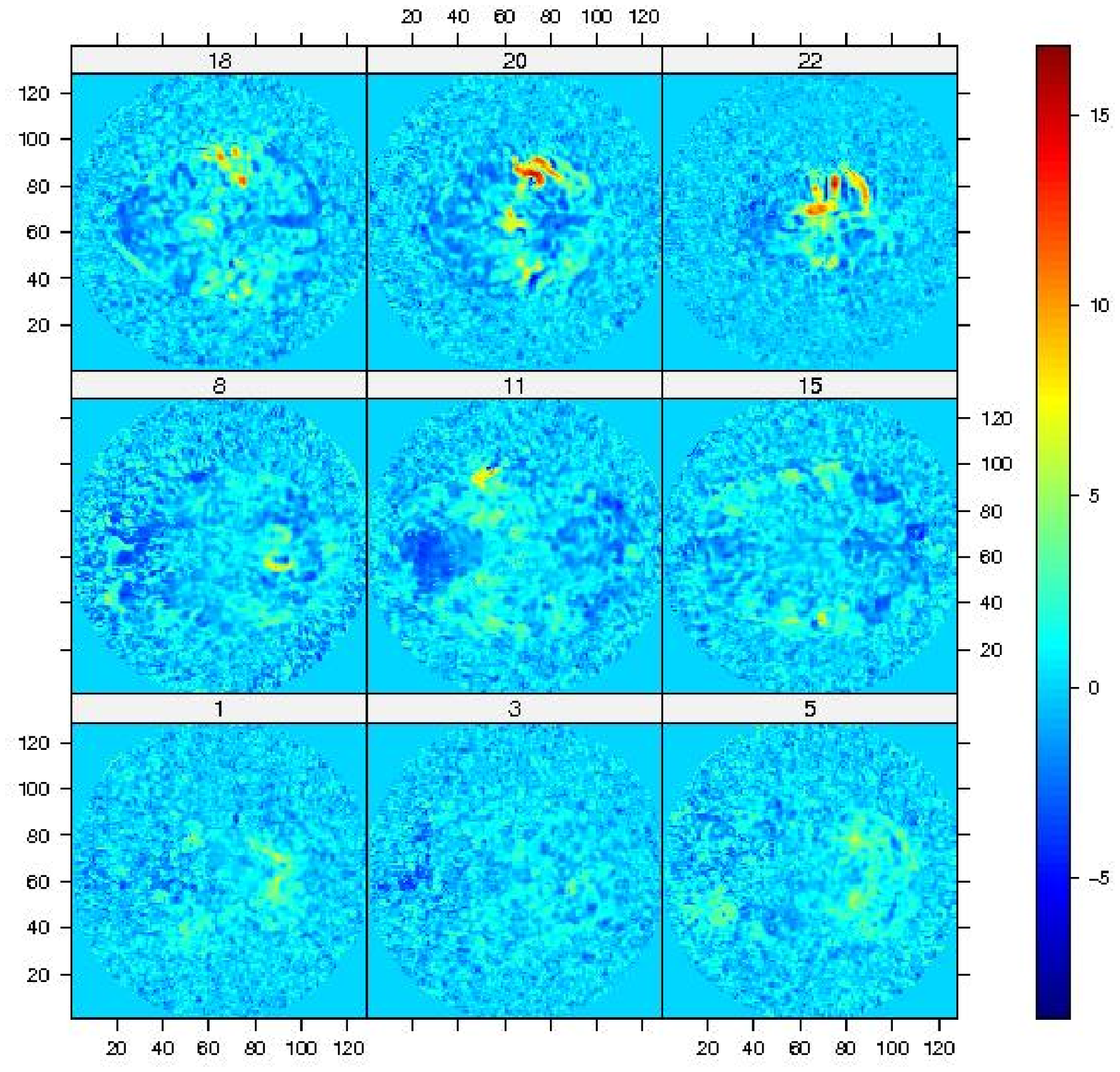} &
\includegraphics[angle=270, width=3.2in]{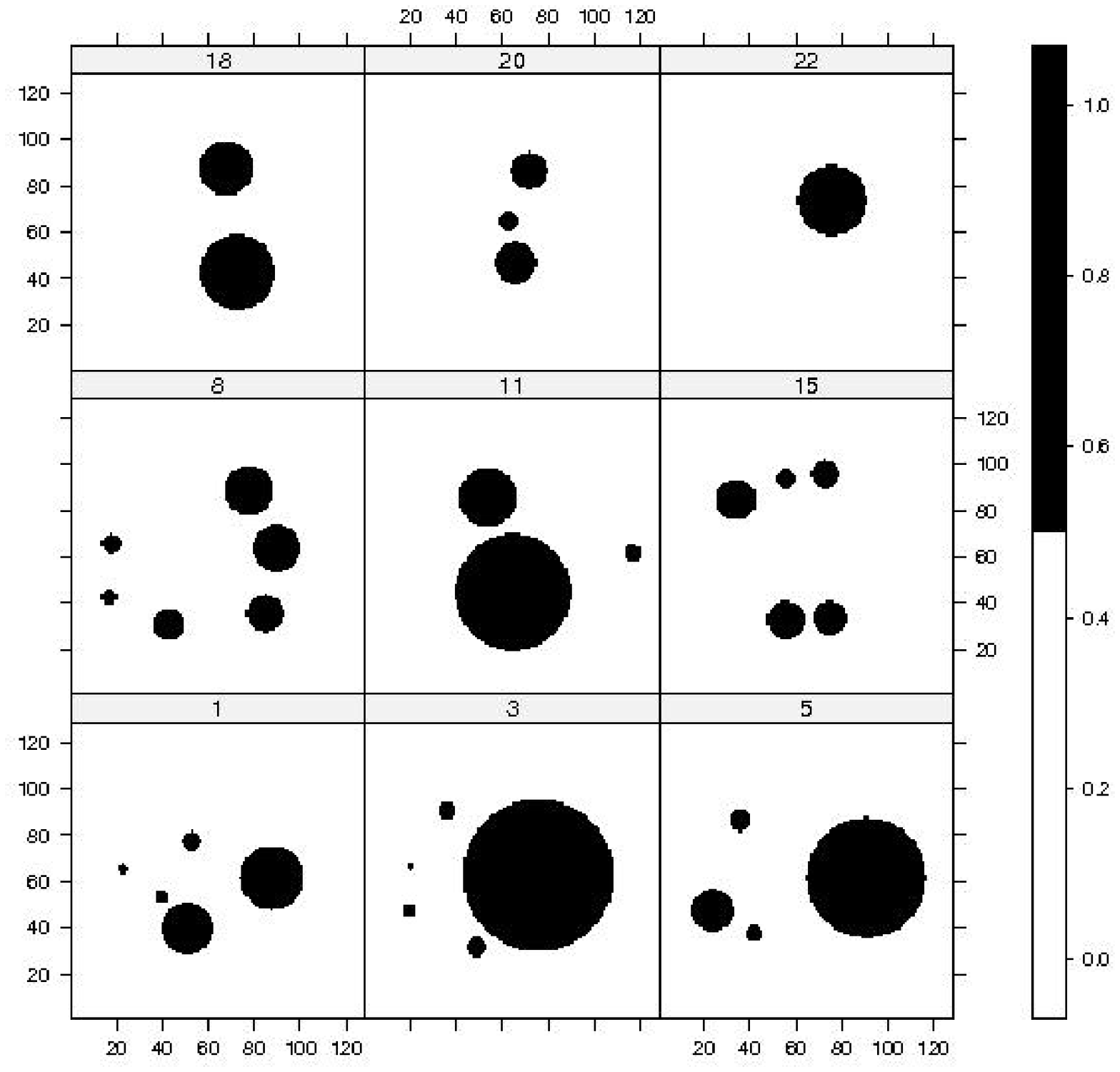} \\
(c) Marginal Detection with FDR control & (d) MCD method \\
\includegraphics[angle=270, width=3.2in]{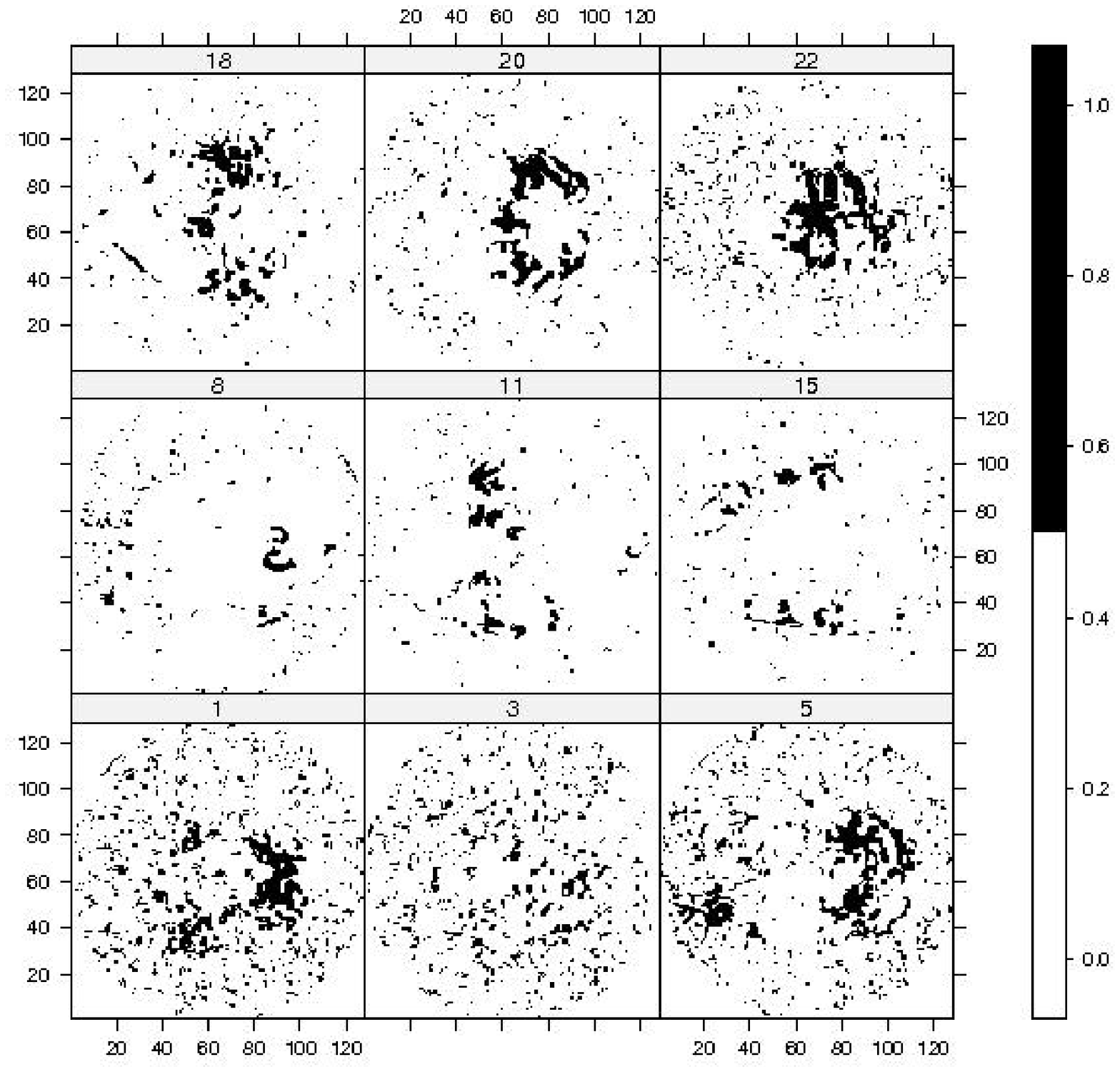} &
\includegraphics[angle=270, width=3.2in]{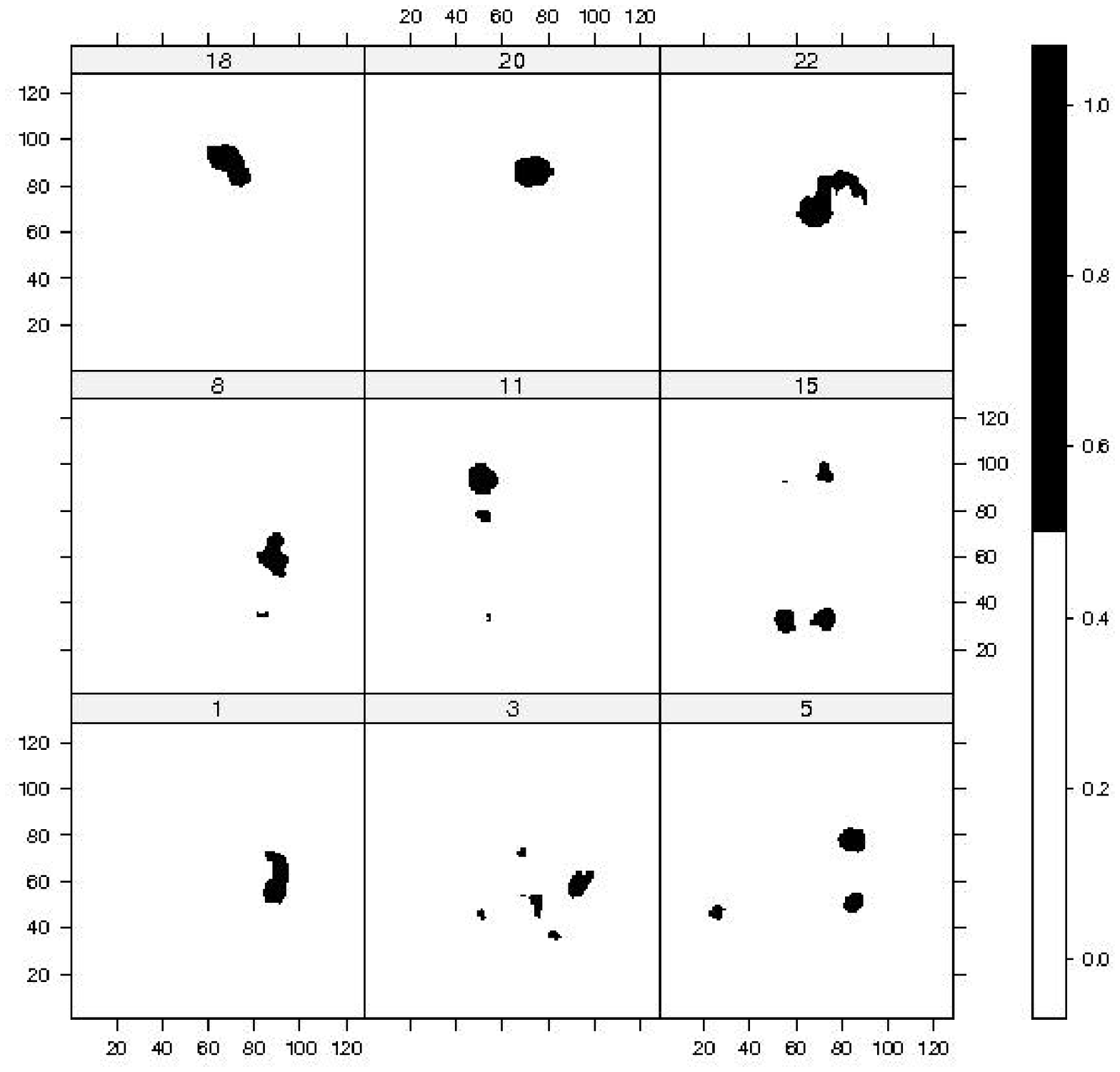}
\end{tabular}
\end{center}
\caption{The detection result for an fMRI data set. The selected 9 slices of image are shown in panel (a). Panel (b) shows the detection result by
the spatial scan method. It identify some interesting regions, but their shapes are regular. Besides, there are some tiny regions outside the big dots, which
may correspond to false discovery. Panel (c) shows marginal detection method with FRD control level at 0.20. It also highlights some regions in each
plot, but these identified spots do not naturally form (continuous) region. Panel (d) provided the detection result by the MCD method. It highlights
several irregular shaped regions in each image.} \label{fMRIFig}
\end{figure*}

\section{Examples: functional MRI data} \label{examplesec}
In this section we apply the three aforementioned methods to a real fMRI data to illustrate their differences in real data application. The fMRI data was firstly analyzed in \cite{Maitra:Roys:Gullapalli:2002, Maitra:2009, Maitra:2010}.

The top left panel in Figure \ref{fMRIFig} shows 9 slices of the fMRI images in a total of 22 slices. Each individual image has 128$\times$128 pixels.
All 9 images show activities in different regions, by showing hot colors (e.g. yellow or red colors which are positive values) and cool colors (dark
blue shows negative values). The response is a transformation of $p$-values from a previous study, and we are only interested in detection of the
positive regions. By visual inspection of the raw data, it appears that there are some (positive) activities at the following regions: central right (slice 1), central left and central right (slice 5),
top left (slice 11), bottom central (slice 15), top central (slice 18), central middle or top central (slice 20), and central part (slice 22). The
shapes of these regions seem to be very irregular.

The bottom left panel shows the single scale detection method with controlling false discovery rate (FDR) (20\% false discovery rate). They highlight
the regions that have been mentioned in the previous paragraph, but there are also many more outside spots, which is likely to be false positives.
In addition, these candidate regions do not naturally form clusters in the figures, which makes it hard to interpret.

The top right panel shows the analysis results by the spatial Scan method. The method shows several important clusters in the plot, such as the clusters
mentioned in previous paragraph. However, these regions are identified as a round shape as suggested by the spatial scan method, and do not reveal any
useful shape information which is important for the analysis of fMRI data.

The bottom right panel shows the results generated by the MCD method. Within each slice, we identified several clusters, which are irregularly shaped.
These regions form natural clusters, and also validate the initial visual impression. These results show that our method is suitable for detection
of irregular shaped spatial signals.

\section{Discussion} \label{discusssec}
In this paper, we proposed a multiresolution cluster detection method, which is shown to have better performance empirically compared to single scale
testing methods controlling for FDR and spatial scan statistics for detecting spatial clusters with irregular shape. We conjecture that the
power of MCD method is larger than methods based on single scale (e.g. the individual test combined with multiple comparison), the theoretical proof of
which will be a future research topic. An early study of multiresolution method in one dimensional situation \citep{zhang:zhu:marron:2007} provides a
foundation for this type of research. Another issue that is worth investigating is the selection of number of scales and the largest scale for the MCD
method, which we did not fully address in this paper.


\section*{Appendix: Derivation of the propositions}
The test statistics can be derived relatively straightforward as the follows:

\subsection*{Derivation of the test statistic under Binomial assumption}
The likelihood function can be calculate by using the following relation:
\begin{align}
&P(X^1=x_1, X^2=x_2, \cdots, X^k=x_k) \notag \\
&=P(X^1=x_1, X^2-X^1=x_2-x_1,
\cdots, \label{calculationformula}\\
&\qquad X^k-X^{(k-1)}=x_k-x_{k-1}) \notag
\end{align}
Note that $\{X^i-X^{(i-1)}\}$ are independent to each other.

Under the null hypothesis, $X^i-X^{(i-1)} \sim Bin(N^i-N^{(i-1)}, p_0)$. The distribution of the alternative is much more complicated than this. Note
that typically $p_0$ and $p_1$ are unknown. We will use the following methods to estimate $p_0$.

Let $\widehat{p}_{ij}= \frac{X_{ij}+1}{N_{ij}+2}$.  Then we have
\[
\widehat{p}_0= \mathrm{Median}(\widehat{p}_{ij}),
\]
which gives a robust estimate of $\widehat{p}_0$. Thus the likelihood under the null is
\begin{align*}
&P(X^1=x_1, X^2=x_2, \cdots, X^k=x_k)\\
&=P(X^1=x_1, X^2-X^1=x_2-x_1,
\cdots, \\
&\qquad X^k-X^{(k-1)}=x_k-x_{k-1}) \\
&=P(X^1=x_1)P(X^2-X^1=x_2-x_1)\cdots \\
&\qquad P(X^k-X^{k-1}=x_k-x_{k-1}) \\
&={N^1 \choose x_1}\widehat{p}_0^{x_1} (1-\widehat{p}_0)^{(N^1-x_1)}{N^2-N^1 \choose x_2-x_1}\widehat{p}_0^{x_2-x_1} \\
&\qquad (1-\widehat{p}_0)^{((N^2-N^1)-(x_2-x_1))} \cdots {N^k-N^{k-1} \choose x_k-x_{k-1}}\widehat{p}_0^{x_k-x_{k-1}}\\
&\qquad (1-\widehat{p}_0)^{((N^k-N^{k-1})-(x_k-x_{k-1}))}.
\end{align*}
The maximum likelihood for the entire parameter space is quite challenging to be calculated. We may further introduce the following assumptions to
simplify the calculation: For each region among the following $k$: $\mathcal{D}_1$, $\mathcal{D}_2-\mathcal{D}_1$, $\cdots$, and
$\mathcal{D}_k-\mathcal{D}_{k-1}$; we assume that the probability of successes are the same. Thus, we could have the following estimate of the
probability of successes:
\begin{align*}
\widehat{p}_1 &= \max (\mathop{\mathrm{Median}}_{\mathcal{D}_1} (\widehat{p}_{ij}), \widehat{p}_0), \\
\widehat{p}_r &= \max (\mathop{\mathrm{Median}}_{\mathcal{D}_r-\mathcal{D}_{r-1}} (\widehat{p}_{ij}), \widehat{p}_0), \quad \forall r\geq 2.
\end{align*}
Based on this, the likelihood for the entire parameter space is
\begin{align*}
&P(X^1=x_1, X^2=x_2, \cdots, X^k=x_k)\\
&=P(X^1=x_1)P(X^2-X^1=x_2-x_1)\cdots \\
&\qquad P(X^k-X^{k-1}=x_k-x_{k-1}) \\
&={N^1 \choose x_1}\widehat{p}_1^{x_1} (1-\widehat{p}_1)^{(N^1-x_1)}{N^2-N^1 \choose x_2-x_1}\widehat{p}_2^{x_2-x_1} \\
&\qquad (1-\widehat{p}_2)^{((N^2-N^1)-(x_2-x_1))} \cdots {N^k-N^{k-1} \choose x_k-x_{k-1}}\widehat{p}_k^{x_k-x_{k-1}} \\
&\qquad (1-\widehat{p}_k)^{((N^k-N^{k-1})-(x_k-x_{k-1}))}.
\end{align*}

Thus, the likelihood ratio will be
\begin{align*}
\Lambda&=\frac{\sup_{\Theta_0}L(\theta; x)}{\sup_{\Theta}L(\theta; x)}\\
&=\left(\frac{\widehat{p}_0}{\widehat{p}_1}\right)^{x_1}
\left(\frac{1-\widehat{p}_0}{1-\widehat{p}_1}\right)^{(N^1-x_1)}\left(\frac{\widehat{p}_0}{\widehat{p}_2}\right)^{x_2-x_1} \\&\quad
\left(\frac{1-\widehat{p}_0}{1-\widehat{p}_2}\right)^{((N^2-N^1)-(x_2-x_1))} \cdots \left(\frac{\widehat{p}_0}{\widehat{p}_k}\right)^{x_k-x_{k-1}} \\
&\quad \left(\frac{1-\widehat{p}_0}{1-\widehat{p}_k}\right)^{((N^k-N^{k-1})-(x_k-x_{k-1}))}.
\end{align*}
Thus, the typical likelihood ratio statistic $-2\log \Lambda$ can be easily calculated, which leads to the proposition.
\subsection*{Derivation of the test statistic under Poisson assumption}
We will use the same calculation formula in \eqref{calculationformula}
For Poisson case, under the null hypothesis, $X^i-X^{i-1} \sim P((|\mathcal{D}_i|-|\mathcal{D}_{i-1}|)\lambda_0)$ for all $i\geq 2$. We can define
$\widehat{\lambda}_0=\mathrm{Median}_\mathcal{D}(X_{ij})$.
Under the alternative or the entire parameter space, the distribution may be the follows, $X^i-X^{i-1} \sim
P((|\mathcal{D}_i|-|\mathcal{D}_{i-1}|)\lambda_i)$, and
\[
\widehat{\lambda}_i=\max(\frac{x_i-x_{i-1}}{|\mathcal{D}_i|-|\mathcal{D}_{i-1}|}, \widehat{\lambda}_0).
\]
Let $m_k^*=\mathcal{D}_k-\mathcal{D}_{k-1}$, we have, under $H_0$,
\begin{align*}
&P(X^1=x_1, X^2=x_2, \cdots, X^k=x_k) \\
&= P(X^1=x_1, X^2-X^1=x_2-x_1, \cdots, \\
&\qquad X^k-X^{k-1}=x_k-x_{k-1}) \\
&=\frac{\widehat{\lambda}_0^{x_1}}{x_1 !} e^{-\widehat{\lambda}_0} \frac{(m_2\widehat{\lambda}_0)^{x_2-x_1}}{(x_2-x_1) !}
e^{-m_2\widehat{\lambda}_0}\cdots \frac{(m_k\widehat{\lambda}_0)^{x_k-x_{k-1}}}{(x_k-x_{k-1}) !} e^{-m_k\widehat{\lambda}_0};
\end{align*}
and under $H_1$,
\begin{align*}
&P(X^1=x_1, X^2=x_2, \cdots, X^k=x_k) \\
&= P(X^1=x_1, X^2-X^1=x_2-x_1, \cdots, \\
&\qquad X^k-X^{k-1}=x_k-x_{k-1}) \\
&=\frac{\widehat{\lambda}_1^{x_1}}{x_1 !} e^{-\widehat{\lambda}_1} \frac{(m_2\widehat{\lambda}_2)^{x_2-x_1}}{(x_2-x_1) !}
e^{-m_2\widehat{\lambda}_2}\cdots \frac{(m_k\widehat{\lambda}_k)^{x_k-x_{k-1}}}{(x_k-x_{k-1}) !} e^{-m_k\widehat{\lambda}_k}.
\end{align*}

The likelihood ratio leads to
\begin{align*}
\Lambda&=\frac{\frac{\widehat{\lambda}_0^{x_1}}{x_1 !} e^{-\widehat{\lambda}_0} \frac{(m_2\widehat{\lambda}_0)^{x_2-x_1}}{(x_2-x_1) !}
e^{-m_2\widehat{\lambda}_0}\cdots \frac{(m_k\widehat{\lambda}_0)^{x_k-x_{k-1}}}{(x_k-x_{k-1}) !} e^{-m_k\widehat{\lambda}_0}}{
\frac{\widehat{\lambda}_1^{x_1}}{x_1 !} e^{-\widehat{\lambda}_1} \frac{(m_2\widehat{\lambda}_2)^{x_2-x_1}}{(x_2-x_1) !}
e^{-m_2\widehat{\lambda}_2}\cdots \frac{(m_k\widehat{\lambda}_k)^{x_k-x_{k-1}}}{(x_k-x_{k-1}) !} e^{-m_k\widehat{\lambda}_k}} \\
&=\left(\frac{\widehat{\lambda}_0}{\widehat{\lambda}_1}\right)^{x_1}\exp(\widehat{\lambda}_1-\widehat{\lambda}_0)\left(\frac{\widehat{\lambda}_0}{\widehat{\lambda}_2}\right)^{x_2-x_1}\exp(m_2(\widehat{\lambda}_2-\widehat{\lambda}_0))
\\
&\qquad \cdots \left(\frac{\widehat{\lambda}_0}{\widehat{\lambda}_k}\right)^{x_k-x_{k-1}}\exp(m_k(\widehat{\lambda}_k-\widehat{\lambda}_0))
\end{align*}
This leads to the proposition

\subsection*{Derivation of the test statistic under Normal assumption}

Under the null hypothesis, we know that $X^i-X^{i-1} \sim N(m_i\mu_0, m_i\sigma^2)$ for all $i\geq 2$. We can define
$\widehat{\mu}_0=\mathrm{Median}_\mathcal{D}(X_{ij})$.

Under the alternative or the entire parameter space, the distribution may be the follows, $X^i-X^{i-1} \sim N(m_i\mu_i, m_i\sigma^2)$, and
\[
\widehat{\mu}_i=\max(\frac{x_i-x_{i-1}}{|\mathcal{D}_i|-|\mathcal{D}_{i-1}|}, \widehat{\mu}_0).
\]

This leads to, under $H_0$,
\begin{align*}
&P(X^1=x_1, X^2=x_2, \cdots, X^k=x_k) \\
& = P(X^1=x_1, X^2-X^1=x_2-x_1, \cdots, \\
&\qquad X^k-X^{k-1}=x_k-x_{k-1})\\
&=\frac{1}{\sqrt{2\pi}\sigma}\exp\left\{\frac{-(x_1-\widehat{\mu}_0)^2}{2\sigma^2}\right\}
\frac{1}{\sqrt{2\pi m_2}\sigma}\\
&\qquad \exp\left\{\frac{-((x_2-x_1)-m_2\widehat{\mu}_0)^2}{2m_2\sigma^2}\right\}\\
& \qquad \cdots \frac{1}{\sqrt{2\pi m_k}\sigma}\exp\left\{\frac{-((x_k-x_{k-1})-m_k\widehat{\mu}_0)^2}{2m_k\sigma^2}\right\};
\end{align*}
and, under $H_1$,
\begin{align*}
&P(X^1=x_1, X^2=x_2, \cdots, X^k=x_k)\\
& = P(X^1=x_1, X^2-X^1=x_2-x_1, \cdots, \\
&\qquad X^k-X^{k-1}=x_k-x_{k-1})\\
&=\frac{1}{\sqrt{2\pi}\sigma}\exp\left\{\frac{-(x_1-\widehat{\mu}_1)^2}{2\sigma^2}\right\}
\frac{1}{\sqrt{2\pi m_2}\sigma}\\
&\qquad \exp\left\{\frac{-((x_2-x_1)-m_2\widehat{\mu}_2)^2}{2m_2\sigma^2}\right\}\\
& \qquad \cdots \frac{1}{\sqrt{2\pi m_k}\sigma}\exp\left\{\frac{-((x_k-x_{k-1})-m_k\widehat{\mu}_k)^2}{2m_k\sigma^2}\right\};
\end{align*}
This leads to
\begin{align*}
\Lambda &=\exp\left\{\frac{1}{2\sigma^2}[(x_1-\widehat{\mu}_1)^2-(x_1-\widehat{\mu}_0)^2]\right.\\
&\qquad
+\frac{1}{2m_2\sigma^2}[((x_2-x_1)-m_2\widehat{\mu}_2)^2-((x_2-x_1)-m_2\widehat{\mu}_0)^2]\\
&\qquad+\cdots+\frac{1}{2m_k\sigma^2}[((x_{k}-x_{k-1})-m_k\widehat{\mu}_k)^2\\
&\qquad -\left.((x_k-x_{k-1})-m_k\widehat{\mu}_0)^2]\right\}\\
&=\exp\left\{\frac{1}{2\sigma^2}[2x_1(\widehat{\mu}_0-\widehat{\mu}_1)+\widehat{\mu}_1^2-\widehat{\mu}_0^2]\right.\\
&\qquad +\frac{1}{2m_2\sigma^2}[2m_2(x_2-x_1)(\widehat{\mu}_0-\widehat{\mu}_2)+m_2^2(\widehat{\mu}_2^2-\widehat{\mu}_0^2)] \\
&\qquad+\cdots+\frac{1}{2m_k\sigma^2}[2m_k(x_k-x_{k-1})(\widehat{\mu}_0-\widehat{\mu}_k)\\
&\qquad+\left.m_k^2(\widehat{\mu}_{k}^2-\widehat{\mu}_0^2)]\right\}.
\end{align*}
This leads to the proposition

\subsection*{Derivation of Theorem \ref{AverageT}}
Let $Y_0$ be the measurement at the specific location, $Y_1$, $Y_2$, $Y_3$, and $Y_4$ are the four measurements collected in $\mathcal{D}_1$. Note that $Y_i \sim N(\mu, 1)$. If $Y_i$ is in the noise region, $Y_i \sim N(0, 1)$, and if $Y_i$ is in the signal region, $Y_i \sim N(\delta, 1)$.

Use the same notation, we have $X_1=Y_0$, and $X_2=\sum_{i=0}^4 Y_i$. The test statistic is defined as
\begin{align*}
-2\log \Lambda &= 2X_1(\widehat{\mu}_1-\widehat{\mu}_0)+(\widehat{\mu}_0^2-\widehat{\mu}_1^2)  \\
&\quad+2(X_2-X_1)(\widehat{\mu}_2-\widehat{\mu}_0)+m_2(\widehat{\mu}_0^2-\widehat{\mu}_2^2)
\end{align*}
It is straightforward to have
\[
\widehat{\mu}_0=\frac{1}{N}\sum_{i=0}^{N-1}Y_i, \quad
\widehat{\mu}_1=Y_0, \quad
\widehat{\mu}_2=\frac{1}{4}\sum_{i=1}^4Y_i,
\]
and $m_2=4$.

Note that $\widehat{\mu}_0 \sim N(n_s\delta/N, 1/N)$. Similarly, we can derive the distribution of $\widehat{\mu}_1$ and $\widehat{\mu}_2$ as follows.
\begin{enumerate}
\item If all $Y_0$, $Y_1$, $\cdots$, $Y_4$ are in the noise region, $\widehat{\mu}_1 \sim N(0, 1)$, and $\widehat{\mu}_2 \sim N(0, 1/4)$.
\item If $Y_0$ is a noise point, but at the boundary as shown in Figure 2 (a) and (b), then
    $\widehat{\mu}_1\sim N(0, 1)$, and $\widehat{\mu}_2 \sim N(2\delta/4, 1/4)$ (case (a)) or $\widehat{\mu}_2\sim N(\delta/4, 1/4)$ (case (b)).
\item If $Y_0$ is a signal point, but at the boundary as shown in Figure 2 (c) and (d), then
    $\widehat{\mu}_1\sim N(\delta, 1)$, and $\widehat{\mu}_2 \sim N(2\delta/4, 1/4)$ (case (c)) or $\widehat{\mu}_2\sim N(3\delta/4, 1/4)$ (case (d)).
\item If $Y_0$, $Y_1$, $\cdots$, $Y_4$ are all in the signal region, $\widehat{\mu}_1 \sim N(\delta, 1)$, and $\widehat{\mu}_2 \sim N(\delta, 1/4)$.
\end{enumerate}

Note that the test statistic, $T$ can be written as
\begin{align*}
T &= 2Y_0\left(Y_0-\frac{1}{N}\sum_{i=0}^{N-1}Y_i\right)+\left(\left(\frac{1}{N}\sum_{i=0}^{N-1}Y_i\right)^2-Y_0^2\right)  \\
&\quad+2 \sum_{i=1}^4Y_i\left(\frac{1}{4}\sum_{i=1}^4Y_i-\frac{1}{N}\sum_{i=0}^{N-1}Y_i\right)\\
&\quad+4\left(\left(\frac{1}{N}\sum_{i=0}^{N-1}Y_i\right)^2-\left(\frac{1}{4}\sum_{i=1}^4Y_i\right)^2\right)
\end{align*}

For each individual $T$, it can be shown that when $N\to \infty$, we have
\[
T\stackrel{d}{\to} \widehat{\mu}_1^2+4\widehat{\mu}_2^2
\]

By using the above consideration on whether $Y_0$ is in the noise region, boundary, and signal region, we have the following results
\begin{enumerate}
\item When all $Y_0$, $Y_1$, $\cdots$, $Y_4$ are in the noise region, $\widehat{\mu}_1^2 \sim \chi^2_1(0)$, $(2\widehat{\mu}_2)^2 \sim \chi^2_1(0)$, and thus $T \stackrel{d}{\to} \chi^2_2(0)$.
\item When $Y_0$ is in noise region, and as shown in Figure 2 (a), $\widehat{\mu}_1^2 \sim \chi^2_1(0)$, $2\widehat{\mu}_2 \sim N(\delta, 1)$, and thus $(2\widehat{\mu}_2)^2 \sim \chi^2_1(\delta^2)$. This leads to $T \stackrel{d}{\to} \chi^2_2(\delta^2)$.
\item When $Y_0$ is in noise region, and as shown in Figure 2 (b), $\widehat{\mu}_1^2 \sim \chi^2_1(0)$, $2\widehat{\mu}_2 \sim N(\delta/2, 1)$, and thus $(2\widehat{\mu}_2)^2 \sim \chi^2_1(\delta^2/4)$. This leads to $T \stackrel{d}{\to} \chi^2_2(\delta^2/4)$.
\item When $Y_0$ is in signal region, and as shown in Figure 2 (c), $\widehat{\mu}_1^2 \sim \chi^2_1(\delta^2)$, $2\widehat{\mu}_2 \sim N(\delta, 1)$, and thus $(2\widehat{\mu}_2)^2 \sim \chi^2_1(\delta^2)$. This leads to $T \stackrel{d}{\to} \chi^2_2(2\delta^2)$.
\item When $Y_0$ is in signal region, and as shown in Figure 2 (d), $\widehat{\mu}_1^2 \sim \chi^2_1(\delta^2)$, $2\widehat{\mu}_2 \sim N(3\delta/2, 1)$, and thus $(2\widehat{\mu}_2)^2 \sim \chi^2_1(9\delta^2/4)$. This leads to $T \stackrel{d}{\to} \chi^2_2(13\delta^2/4)$.
\item When $Y_0$, $Y_1$, $\cdots$, $Y_4$ are in the signal region, $\widehat{\mu}_1^2 \sim \chi^2_1(\delta^2)$, $2\widehat{\mu}_2 \sim N(2\delta, 1)$, and thus $(2\widehat{\mu}_2)^2 \sim \chi^2_1(4\delta^2)$. This leads to $T \stackrel{d}{\to} \chi^2_2(5\delta^2)$.
\end{enumerate}

Note that $\widehat{\mu}_1$ and $\widehat{\mu}_2$ are independent to each other. Thus, we will have following relations:

\[
E(T)=\left\{
\begin{array}{ll}
2 & Y_0 \textnormal{ is an inner point in the noise region} \\
f(\delta) & Y_0 \textnormal{ is an boundary point} \\
g(\delta) & Y_0 \textnormal{ is an inner point in the signal region} \\
\end{array}
\right.,
\]
where $2+\delta^2/4 \leq f(\delta) \leq 2+13\delta^2/4$, and $g(\delta)=2+5\delta^2$. It is straightforward that, if $\delta>0$,
\[
2<f(\delta)<g(\delta).
\]
Similarly, we have

\[
V(T)=\left\{
\begin{array}{ll}
4 & Y_0 \textnormal{ is an inner point in the noise region} \\
v_1(\delta) & Y_0 \textnormal{ is an boundary point} \\
v_2(\delta) & Y_0 \textnormal{ is an inner point in the signal region} \\
\end{array}
\right.,
\]
where $2(2+\delta^2/2) \leq v_1(\delta) \leq 2(2+13\delta^2/2)$, and $v_2(\delta)=2(2+10\delta^2)$.

Let us assume that we have $n_n$ inner points in the noise region, $n_b$ boundary points, and $n_s$ inner points in the signal region. The average $T(y_0)$ will have
\begin{align*}
\frac{1}{n_n}\sum_{Y_0 \in \mathcal{N}} T(Y_0) & \stackrel{d}{\to} N\left(2, \frac{4}{n_n}\right)\\
\frac{1}{n_b}\sum_{Y_0 \in \mathcal{B}} T(Y_0) & \stackrel{d}{\to} N\left(f(\delta), \frac{v_1(\delta)}{n_b}\right)\\
\frac{1}{n_s}\sum_{Y_0 \in \mathcal{S}} T(Y_0) & \stackrel{d}{\to} N\left(g(\delta), \frac{v_2(\delta)}{n_s}\right)
\end{align*}
Thus, if $n_n$, $n_b$, and $n_s \to \infty$, we will have stochastically
\[
P\left(\frac{1}{n_n}\sum_{Y_0 \in \mathcal{N}} T(Y_0) < \frac{1}{n_b}\sum_{Y_0 \in \mathcal{B}} T(Y_0) < \frac{1}{n_s}\sum_{Y_0 \in \mathcal{S}} T(Y_0)\right)\to 1.
\]

\subsection*{Derivation of Theorem \ref{AverageV}}

Let us use similar notation as above. Let $\mathcal{B}$ be the boundary point, which is defined as at least one of its neighbor 4 points is from another group. $\mathcal{B}^c$ be non-boundary point. The local variability is defined as $\frac{1}{4}\sum_{i=0}^4(Y_i-\overline{Y})^2$. Let us focus on $\widetilde{V}(Y)=\sum_{i=0}^4(Y_i-\overline{Y})^2$. It is straightforward to show that if $Y\in \mathcal{B}^c$, $\widetilde{V}(Y) \sim \chi^2_4(0)$. Similarly as derivation in the former section, let $Y_i \sim N(\mu_i, 1)$. If $Y_0\in \mathcal{B}$, we have $\sum Y_i^2 \sim \chi^2_5(\sum \mu_i^2)$. It can be shown that $\mu_i^2=k \delta^2$, where $k$ is the number of signal points in the local region (note that, $k=1, 2, 3, 4$). Let $\overline{Y}=\frac{1}{5}\sum_{i=0}^4Y_i$, It is straightforward that $\overline{Y} \sim N(k\delta/5, 1/5)$, or $\sqrt{5}\overline{Y}\sim N(k\delta/\sqrt{5}, 1)$. And thus
\[
\widetilde{V}(Y) \sim \chi^2_4(C),
\]
where
\[
C=k\delta^2-k^2\delta^2/5, \quad k=1, 2, 3, 4.
\]
Then, $E(\widetilde{V}(Y))=4+C$, and $\mathrm{Var}(\widetilde{V}(Y))=8+4C$, if $Y\in \mathcal{B}$. Similarly, $E(\widetilde{V}(Y))=4$, and $\mathrm{Var}(\widetilde{V}(Y))=8$, when $Y\in \mathcal{B}^c$. By the central limit theorem, we have
\begin{align*}
\mathop{\mathrm{Ave}}_{Y\in \mathcal{B}}\widetilde{V}(Y) & \stackrel{d}{\to} N\left(4+C, \frac{8+4C}{n_b}\right) \\
\mathop{\mathrm{Ave}}_{Y\in \mathcal{B}^c}\widetilde{V}(Y) & \stackrel{d}{\to} N\left(4, \frac{8}{N-n_b}\right) \\
\end{align*}
And thus
\[
\mathop{\mathrm{Ave}}_{Y\in \mathcal{B}}\widetilde{V}(Y)-\mathop{\mathrm{Ave}}_{Y\in \mathcal{B}^c}\widetilde{V}(Y) \stackrel{d}{\to} N\left(C, \frac{8+4C}{n_b}+\frac{8}{N-n_b}\right)
\]
If we want to $P(\mathop{\mathrm{Ave}}_{Y\in \mathcal{B}}\widetilde{V}(Y)-\mathop{\mathrm{Ave}}_{Y\in \mathcal{B}^c}\widetilde{V}(Y) >0) \to 1$, we need to have
\[
\Phi\left(a\right) \to 1.
\]
where
\[
a=\sqrt{\frac{CNp_B(1-p_B)}{8+4(1-p_B)}}.
\]
Here $p_B$ be the proportion of boundary point in the image. So if $\frac{\delta^2Np_B(1-p_B)}{8+4(1-p_B)} \to \infty$, then $a\to \infty$. This leads to $P(\mathrm{Ave}_{Y\in \mathcal{B}}\widetilde{V}(Y)-\mathrm{Ave}_{Y\in \mathcal{B}^c}\widetilde{V}(Y) >0) \to 1$. And then $P(\mathrm{Ave}_{Y\in \mathcal{B}}{V}(Y)-\mathrm{Ave}_{Y\in \mathcal{B}^c}{V}(Y) >0) \to 1$.

\bibliography{mcdpaper-sii}

\begin{thebibliography}{29}
\newcommand{\enquote}[1]{``#1''}
\expandafter\ifx\csname natexlab\endcsname\relax\def\natexlab#1{#1}\fi

\bibitem[{Baker(2004)}]{Bake:modi:2004}
Baker, R.~D. (2004), \enquote{A Modified {K}nox Test of Space-time Clustering,}
  \textit{Journal of Applied Statistics}, 31, 457--463.

\bibitem[{Benjamini and Hochberg(1995)}]{benjamini1995controlling}
Benjamini, Y. and Hochberg, Y. (1995), \enquote{Controlling the False Discovery
  Rate: A Practical and Powerful Approach to Multiple Testing,} \textit{Journal
  of the Royal Statistical Society. Series B (Methodological)}, 57, 289--300.

\bibitem[{Chaudhuri and Marron(1999)}]{chaudhuri1999sizer}
Chaudhuri, P. and Marron, J. (1999), \enquote{SiZer for Exploration of
  Structures in Curves,} \textit{Journal of the American Statistical
  Association}, 94, 807--823.

\bibitem[{Chaudhuri and Marron(2000)}]{chaudhuri2000scale}
--- (2000), \enquote{Scale space view of curve estimation,} \textit{The Annals
  of Statistics}, 28, 408--428.

\bibitem[{Costa and Kulldorff(2009)}]{costa2009applications}
Costa, M. and Kulldorff, M. (2009), \enquote{Applications of Spatial Scan
  Statistics: A Review,} \textit{Scan Statistics}, 129--152.

\bibitem[{Diggle et~al.(2005)Diggle, Rowlingson, and Su}]{Diggle:et.al:2005}
Diggle, P., Rowlingson, B., and Su, T.-l. (2005), \enquote{Point process
  methodology for on-line spatio-temporal disease surveillance,}
  \textit{Environmetrics}, 16, 423--434.

\bibitem[{Glaz and Balakrishnan(1999)}]{glaz1999scan}
Glaz, J. and Balakrishnan, N. (1999), \enquote{Scan statistics and
  applications,} .

\bibitem[{Glaz et~al.(2001)Glaz, Naus, and Wallenstein}]{glaz2001scan}
Glaz, J., Naus, J., and Wallenstein, S. (2001), \textit{Scan statistics},
  Springer Verlag.

\bibitem[{Glaz et~al.(2009)Glaz, Pozdnyakov, and Wallenstein}]{glaz2009scan}
Glaz, J., Pozdnyakov, V., and Wallenstein, S. (2009), \textit{Scan statistics:
  methods and applications}, Birkhauser.

\bibitem[{Kulldorff(1999)}]{kulldorff1999spatial}
Kulldorff, M. (1999), \enquote{Spatial scan statistics: models, calculations,
  and applications,} \textit{Scan statistics and applications}, 303--322.

\bibitem[{Kulldorff(2010)}]{kulldorff2010satscan}
--- (2010), \enquote{SaTScan-Software for the spatial, temporal, and space-time
  scan statistics,} \textit{Boston: Harvard Medical School and Harvard Pilgrim
  Health Care}.

\bibitem[{Kulldorff et~al.(2006)Kulldorff, Huang, Pickle, and
  Duczmal}]{kulldorff2006elliptic}
Kulldorff, M., Huang, L., Pickle, L., and Duczmal, L. (2006), \enquote{An
  elliptic spatial scan statistic,} \textit{Statistics in medicine}, 25,
  3929--3943.

\bibitem[{Kulldorff et~al.(2003)Kulldorff, Tango, and
  Park}]{Kull:Tang:Park:powe:2003}
Kulldorff, M., Tango, T., and Park, P.~J. (2003), \enquote{Power Comparisons
  for Disease Clustering Tests,} \textit{Computational Statistics \& Data
  Analysis}, 42, 665--684.

\bibitem[{Liang et~al.(2009)Liang, Banerjee, and
  Carlin}]{Lian:Bane:Carl:baye:2009}
Liang, S., Banerjee, S., and Carlin, B.~P. (2009), \enquote{Bayesian {W}ombling
  for {S}patial {P}oint {P}rocesses,} \textit{Biometrics}, 65, 1243--1253.

\bibitem[{Lindeberg(1993)}]{lindeberg1993scale}
Lindeberg, T. (1993), \textit{Scale-space theory in computer vision}, Springer.

\bibitem[{Lindeberg(1994)}]{lindeberg1994scale}
--- (1994), \enquote{Scale-space theory: A basic tool for analyzing structures
  at different scales,} \textit{Journal of applied statistics}, 21, 225--270.

\bibitem[{Loh({2011})}]{Loh2011}
Loh, J.~M. ({2011}), \enquote{{K-scan for anomaly detection in disease
  surveillance},} \textit{{ENVIRONMETRICS}}, {22}, {179--191}.

\bibitem[{Loh et~al.(2008)Loh, Lindquist, and Wager}]{Loh:Lind:Wage:resi:2008}
Loh, J.~M., Lindquist, M.~A., and Wager, T.~D. (2008), \enquote{Residual
  Analysis for Detecting Mis-modeling in F{MRI},} \textit{Statistica Sinica},
  18, 1421--1448.

\bibitem[{Maitra(2009)}]{Maitra:2009}
Maitra, R. (2009), \enquote{Assessing Certainty of Activation or Inactivation
  in Test-Retest fMRI Studies,} \textit{Neuroimage}, 47, 88--97, dOI
  information: 10.1016/j.neuroimage.2009.03.073.

\bibitem[{Maitra(2010)}]{Maitra:2010}
--- (2010), \enquote{A Re-defined and Generalized Percent-Overlap-of-Activation
  Measure for Studies of fMRI Reproducibility and its Use in Identifying
  Outlier Activation Maps,} \textit{Neuroimage}, 50, 124--135, dOI information:
  10.1016/j.neuroimage.2009.11.070.

\bibitem[{Maitra et~al.(2002)Maitra, Roys, and
  Gullapalli}]{Maitra:Roys:Gullapalli:2002}
Maitra, R., Roys, S.~R., and Gullapalli, R.~P. (2002), \enquote{Test-Retest
  Reliability Estimation of fMRI Data,} \textit{Magnetic Resonance in
  Medicine}, 48, 62--70.

\bibitem[{Naus(1965{\natexlab{a}})}]{naus1965clustering}
Naus, J. (1965{\natexlab{a}}), \enquote{Clustering of random points in two
  dimensions,} \textit{Biometrika}, 52, 263--267.

\bibitem[{Naus(1965{\natexlab{b}})}]{naus1965distribution}
--- (1965{\natexlab{b}}), \enquote{The distribution of the size of the maximum
  cluster of points on a line,} \textit{Journal of the American Statistical
  Association}, 532--538.

\bibitem[{Storey(2002)}]{storey2002direct}
Storey, J. (2002), \enquote{A direct approach to false discovery rates,}
  \textit{Journal of the Royal Statistical Society: Series B (Statistical
  Methodology)}, 64, 479--498.

\bibitem[{Wheeler({2007})}]{wheeler2007}
Wheeler, D.~C. ({2007}), \enquote{{A comparison of spatial clustering and
  cluster detection techniques for childhood leukemia incidence in Ohio,
  1996-2003},} \textit{{INTERNATIONAL JOURNAL OF HEALTH GEOGRAPHICS}}, {6}.

\bibitem[{Zhang et~al.(2011)Zhang, Fan, and Yu}]{zhang2011multiple}
Zhang, C., Fan, J., and Yu, T. (2011), \enquote{Multiple Testing via $FDR_l$
  For Large-Scale Imaging Data,} \textit{Annals of statistics}, 39, 613--642.

\bibitem[{Zhang et~al.(2008)Zhang, Zhu, Jeffay, Marron, and
  Smith}]{zhang2008multi}
Zhang, L., Zhu, Z., Jeffay, K., Marron, J., and Smith, F. (2008),
  \enquote{Multi-resolution anomaly detection for the internet,} in
  \textit{INFOCOM Workshops 2008, IEEE}, IEEE, pp. 1--6.

\bibitem[{Zhang et~al.(2007)Zhang, Zhu, and Marron}]{zhang:zhu:marron:2007}
Zhang, L., Zhu, Z., and Marron, J.~S. (2007), \enquote{Multiresolution anomaly
  detection method for long range dependent time series,} Technical Report,
  UNC/STOR/07/12.

\bibitem[{Zhang and Lin(2009)}]{Zhan:Lin:clus:2009}
Zhang, T. and Lin, G. (2009), \enquote{Spatial scan statistics in loglinear
  models,} \textit{Computational Statistics and Data Analysis}, 53, 2851--2858.

\end{thebibliography}

\end{document}